\documentclass[preprint,aps,superscriptaddress,longbibliography]{revtex4-2}

\usepackage{amssymb}
\usepackage{amsmath}
\usepackage{amsfonts}
\usepackage{bm}
\usepackage{graphicx}
\usepackage{color}
\usepackage[usenames,dvipsnames,svgnames]{xcolor}
\usepackage{bm}
\usepackage{booktabs}
\usepackage{nicefrac}
\usepackage{threeparttable}
\renewcommand{\arraystretch}{0.7}
\usepackage{upgreek}

\usepackage[displaymath]{lineno}

\usepackage{hyperref}
\hypersetup{
	colorlinks,
	linkcolor={blue},
	citecolor={blue},
	urlcolor={blue}
}

\newcommand{\cebini}{Ce$_3$Bi$_4$Ni$_3$}
\newcommand{\cebipd}{Ce$_3$Bi$_4$Pd$_3$}
\newcommand{\cebipt}{Ce$_3$Bi$_4$Pt$_3$}
\newcommand{\rpx}{$R_{3}Pn_{4}X_{3}$}
\newcommand{\cebix}{Ce$_{3}$Bi$_{4}X_{3}$}

\begin{document}


\title{\cebini\ -- A large hybridization-gap variant of \cebipt}

\author{D.\ M.\ Kirschbaum}
\affiliation{Institute of Solid State Physics, Vienna University of Technology, 1040 Vienna, Austria}
\author{X.\ Yan}
\affiliation{Institute of Solid State Physics, Vienna University of Technology, 1040 Vienna, Austria}
\author{M.\ Waas}
\affiliation{Institute of Solid State Physics, Vienna University of Technology, 1040 Vienna, Austria}
\author{R.\ Svagera}
\affiliation{Institute of Solid State Physics, Vienna University of Technology, 1040 Vienna, Austria}
\author{A.\ Prokofiev}
\affiliation{Institute of Solid State Physics, Vienna University of Technology, 1040 Vienna, Austria}
\author{B.\ St\"{o}ger}
\affiliation{X-ray Center, Vienna University of Technology, 1040 Vienna, Austria}
\author{G.\ Giester}
\affiliation{Institute of Mineralogy and Crystallography, University of Vienna, 1090 Vienna, Austria}
\author{P.\ Rogl}
\affiliation{Institute of Materials Chemistry, University of Vienna, 1090 Vienna, Austria}
\author{D.-G.\ Oprea}
\affiliation{Max Planck Institute for Chemical Physics of Solids, 01187 Dresden, Germany}
\affiliation{Institute for Theoretical Physics, Goethe University Frankfurt, 60438 Frankfurt a.M., Germany}
\author{C.\ Felser}
\affiliation{Max Planck Institute for Chemical Physics of Solids, 01187 Dresden, Germany}
\author{R.\ Valent\'{i}}
\affiliation{Institute for Theoretical Physics, Goethe University Frankfurt, 60438 Frankfurt a.M., Germany}
\author{M.\ G.\ Vergniory}
\affiliation{Max Planck Institute for Chemical Physics of Solids, 01187 Dresden, Germany}
\affiliation{Donostia International Physics Center, 20018 Donostia/San Sebasti\'{a}n, Spain}
\author{J.\ Custers}
\affiliation{Dept.\ of Condensed Matter Physics, Charles University, 121 16 Prague, Czech Republic}
\author{S.\ Paschen}
\email[Corresponding author: ]{paschen@ifp.tuwien.ac.at}
\affiliation{Institute of Solid State Physics, Vienna University of Technology, 1040 Vienna, Austria}
\author{D.\ A.\ Zocco}
\email[Corresponding author: ]{zocco@ifp.tuwien.ac.at}
\affiliation{Institute of Solid State Physics, Vienna University of Technology, 1040 Vienna, Austria}
\date{\today}

\begin{abstract}
The family of cubic noncentrosymmetric 3-4-3 compounds has become a fertile ground for the discovery of novel correlated metallic and insulating phases. Here, we report the synthesis of a new heavy fermion compound, \cebini. It is an isoelectronic analog of the prototypical Kondo insulator \cebipt\ and of the recently discovered Weyl-Kondo semimetal \cebipd. In contrast to the volume-preserving Pt-Pd substitution, structural and chemical analyses reveal a positive chemical pressure effect in \cebini\ relative to its heavier counterparts. Based on the results of electrical resistivity, Hall effect, magnetic susceptibility, and specific heat measurements, we identify an energy gap of 65--70\,meV, about eight times larger than that in \cebipt\ and about 45 times larger than that of the Kondo-insulating background hosting the Weyl nodes in \cebipd. We show that this gap as well as other physical properties do not evolve monotonically with increasing atomic number, i.e., in the sequence \cebini-\cebipd-\cebipt, but instead with increasing partial electronic density of states of the $d$ orbitals at the Fermi energy. This work opens the possibility to investigate the conditions under which topological states develop in this series of strongly correlated 3-4-3 materials.
\end{abstract}

\maketitle

\section{Introduction}\label{intro}
The prospect that strong electron correlations and nontrivial electronic topology could cooperate in the formation of novel quantum states of matter has sparked great interest in recent years. Much attention has been given to heavy fermion materials, with proposals to realize correlated topological states in Kondo insulators and semimetals \cite{dzero16a,li20a,paschen21}. In the family of cubic, noncentrosymmetric Ce-based ``3-4-3'' compounds of space group 220, \cebipt\ is perhaps the best-known case. It was classified early on as a Kondo insulator (KI) \cite{fisk88a,hundley90a,fisk92a}, a material where the Kondo interaction between localized (typically 4$f$) states and conduction electrons (typically from $d$ shells) promotes the formation of a narrow band gap \cite{riseborough00a}. More recently it was found that the isoelectronic and isosize substitution of Pt by Pd consecutively reduces the gap, ultimately giving rise to a novel Weyl-Kondo semimetal (WKSM) state in the end compound \cebipd\ \cite{dzsaber17a,lai18a,dzsaber21a}.

Magnetic field experiments showed that the WKSM (gapless) signatures observed in \cebipd\ can be readily suppressed at a modest field of 9\,T while maintaining the Kondo interaction essentially intact \cite{dzsaber22a,grefe20b,grefe21a}. The background Kondo gap, albeit small, remains finite up to a considerably larger magnetic field, as demonstrated from magnetic torque and susceptibility experiments \cite{dzsaber22a,kushwaha19a}. High pressure studies indicate that a larger gap value similar to that of \cebipt\ is gradually recovered \cite{ajeesh22a}. Yet, a detailed understanding of how the WKSM state is affected by volume and hybridization changes produced by chemical substitutions \cite{dzsaber17a, tomczak20a} or externally applied pressure \cite{ajeesh22a,xu22a,zhang22a} is currently lacking.

Here we extend the previous isoelectronic substitution study (5$d$ to 4$d$) one step further, to the Ni case (3$d$). We present the synthesis of high-quality single crystals of \cebini, a material that has not been synthesized in any form before. Structural analysis reveals that the Ni substitution results in a positive chemical pressure, distinct from the volume-preserving Pt-Pd substitution, which gives new insight into the pressure effects on \cebipd. Electrical resistivity, Hall effect, magnetic susceptibility, and specific heat measurements suggest that \cebini\ is a new heavy fermion compound with a hybridization gap that is large compared to those of the Pd and Pt compounds. We end by discussing the evolution of the characteristic energy scales across the \cebix\ series ($X$ = Ni, Pd, Pt), and relate it to the results of electronic band structure calculations.

\section{Single crystal growth and characterization}\label{crystalgrowth}

Materials with the chemical formula \rpx, where $R$ is a rare earth or actinide element, $Pn$ is a pnictogen (As, Sb, Bi), and $X$ is a transition metal element, are extensively investigated \cite{dwight77a,dwight79a,takabatake90a,wang99a}. Among the Ce compounds, the Kondo insulator \cebipt\ has been known for decades \cite{fisk88a,hundley90a,fisk92a}. While single crystals of the isoelectronic compound \cebipd\ were recently obtained via Bi-flux technique \cite{dzsaber17a}, the synthesis of the Ni phase has remained elusive so far. Our initial attempts to prepare polycrystalline samples confirmed that the \cebini\ phase indeed forms, and encouraged us to attempt single crystal growth using various fluxes. A detailed description of the sample synthesis is presented in Appendix \ref{apx-methods}.

\begin{figure}[h]
\centering
\includegraphics[width=0.6\columnwidth]{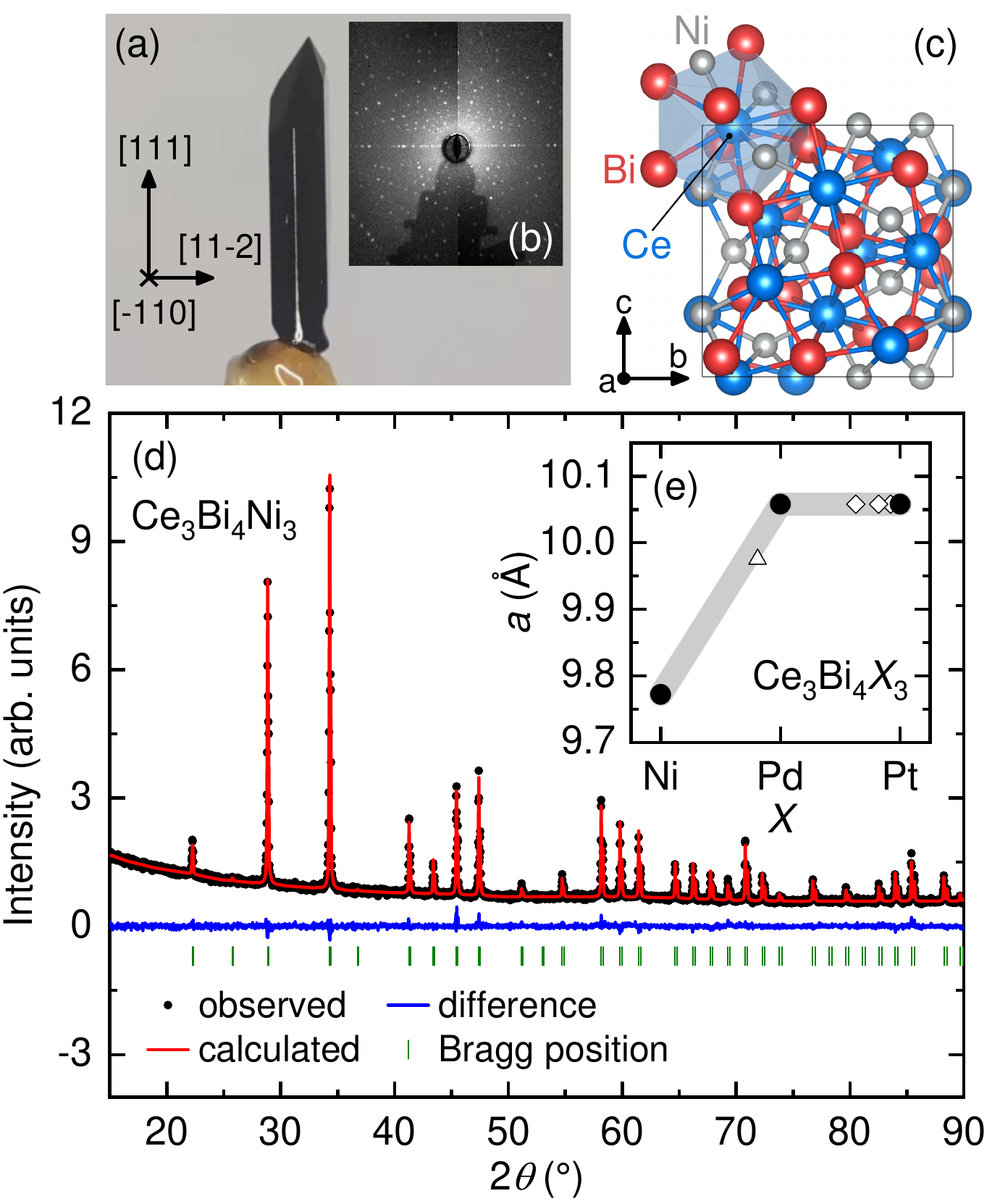}
\caption{\textbf{(a)} Photo of a single crystal of \cebini\ with indicated crystallographic directions obtained from Laue diffraction. Typical sample length is 1.5\,mm. \textbf{(b)} Laue XRD pattern (white spots). The center of the image corresponds to the [-2-1-1] crystallographic direction. \textbf{(c)} Sketch of the unit cell. The polyhedron emphasizes the close environment of a Ce atom. \textbf{(d)} Room-temperature powder XRD pattern of \cebini\ (black dots). Vertical bars indicate the positions of the expected diffraction peaks. The difference (blue line) between the measured data and the refinement (red line) shows no signs of impurity phases. \textbf{(e)} Lattice parameter $a$ across the \cebix\ series ($X$ = Ni, Pd, Pt). Intermediate Pd-Pt values (white diamonds) taken from Ref.\,\cite{dzsaber17a}. The gray shaded line is a guide to the eye.}
\label{fig-powderXRD}
\end{figure}

Single crystals of \cebini\ with faceted surfaces and maximum sizes of $2\times0.5\times0.5$\,mm$^3$ were successfully grown using Pb flux [Fig.\,\hyperref[fig-powderXRD]{\ref{fig-powderXRD}(a)}]. Laue x-ray diffraction (XRD) provides clear patterns, confirming the high quality of the crystals [Fig.\,\hyperref[fig-powderXRD]{\ref{fig-powderXRD}(b)}]. Similarly to the Pt and Pd 3-4-3 compounds, the powder and single-crystal XRD data of \cebini\ can be refined with the cubic Y$_{3}$Sb$_{4}$Au$_{3}$ structure, space group 220, $I\overline{4}$3$d$ \cite{pearson72a,dwight77a}. The unit cell contains four formula units, with 40 atoms forming a noncentrosymmetric structure. Figure \hyperref[fig-powderXRD]{\ref{fig-powderXRD}(c)} displays the crammed unit cell; the local environment of a Ce atom is comprised of eight Bi atoms and four Ni atoms (nearest neighbors, see Appendix \ref{apx-structure}, Tables\,\ref{table1}, \ref{table2}, and \ref{table3}). Rietveld refinements of the powder XRD data of the most stoichiometric samples [Fig.\,\hyperref[fig-powderXRD]{\ref{fig-powderXRD}(d)}] yield an average room-temperature lattice parameter $a = 9.7715(9)$\,\AA\ (see Appendix \ref{apx-stoichiometry}). The unit cell volume of \cebini\ is therefore compressed by $\sim 8$\% with respect to the (essentially isovolume \cite{dzsaber17a}) Pt ($a = 10.051$\,\AA) and Pd ($a = 10.058$\,\AA) compounds [Fig.\,\hyperref[fig-powderXRD]{\ref{fig-powderXRD}(e)}]. These results were confirmed and studied in more detail with single-crystal XRD experiments performed at 100\,K, 200\,K, and 300\,K. A summary of the structural data obtained from a highly stoichiometric sample, including interatomic distances, atomic occupancies, and atomic displacement parameters (ADPs), is presented in Appendix \ref{apx-structure}.

\begin{figure}[h]
\centering
\includegraphics[width=0.6\columnwidth]{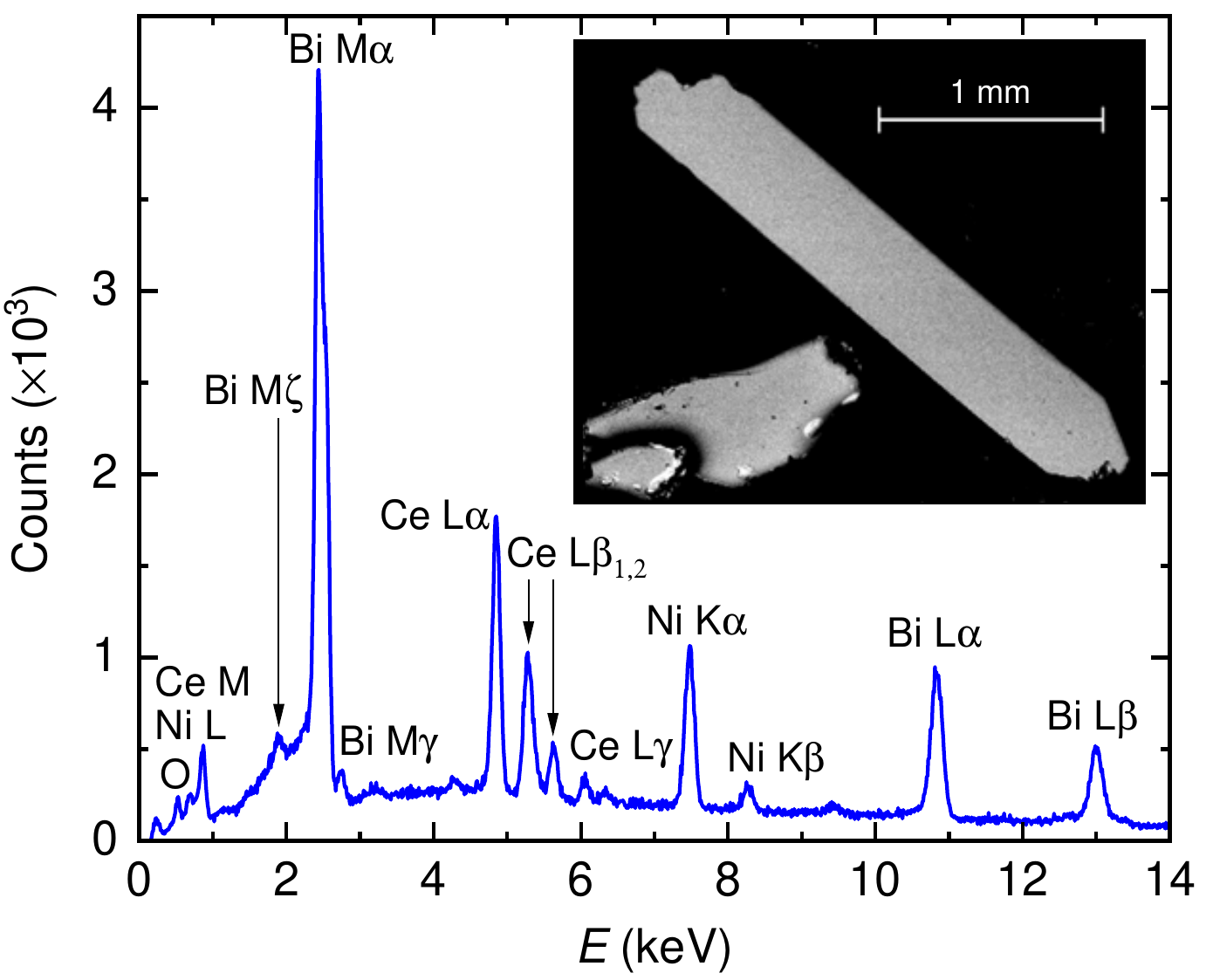}
\caption{EDX spectrum corresponding to one point measurement of a polished \cebini\ crystal (inset, top right). Inset: Backscattered SEM images of as-grown (bottom left) and polished (top right) \cebini\ samples. The lighter spots visible on the surface of the as-grown crystal correspond to Pb, and are absent in the polished sample.}
\label{fig-EDX}
\end{figure}

The stoichiometries of the samples were determined via energy dispersive x-ray spectroscopy (EDX) after each growth campaign. Figure \ref{fig-EDX} displays a representative EDX spectrum collected from the polished surface of a high-quality single crystal. The mean composition Ce$_{3.15(5)}$Bi$_{3.90(9)}$Ni$_{2.94(4)}$ was obtained from spectra taken at six different spots of a sample. Scanning electron microscopy (SEM) images obtained from as-grown single crystals (inset of Fig.\,\ref{fig-EDX}, bottom left) show lighter spots, which could originate from residual flux solidifying after centrifuging. In high-quality single crystals, these inclusions are only present at the surface of the samples, and can therefore be removed by etching with hydrogen peroxide (H$_2$O$_2$) and careful polishing (inset of Fig.\,\ref{fig-EDX}, top right). Spurious superconductivity can therefore be prevented or eliminated, as we confirm by magnetic susceptibility and electrical resistivity measurements. A possible substitution of Bi by Pb in \cebini\ has been ruled out by comparing the EDX spectra of a Pb-flux grown single crystal and that of a Pb-free annealed polycrystalline sample. 

Samples of \cebini\ are sensitive to air and moisture. A distinct degradation of the powder XRD intensity was observed after a 5\,h exposure of the powder to air. The oxygen contamination can also be detected in the EDX spectrum after a short exposure time (small peak in the low energy region of Fig.\,\ref{fig-EDX}).

From a combined analysis of the EDX, powder XRD, and magnetic susceptibility measurements, we can select with good certainty the most stoichiometric, high-quality single crystals. A detailed discussion of this analysis is presented in Appendix \ref{apx-stoichiometry}. In the following sections we present the physical properties of \cebini\ corresponding to samples with Ni content closest to three atoms per mol.

\section{Physical properties}\label{results}

\begin{figure}[h]
\centering
\includegraphics[width=0.6\columnwidth]{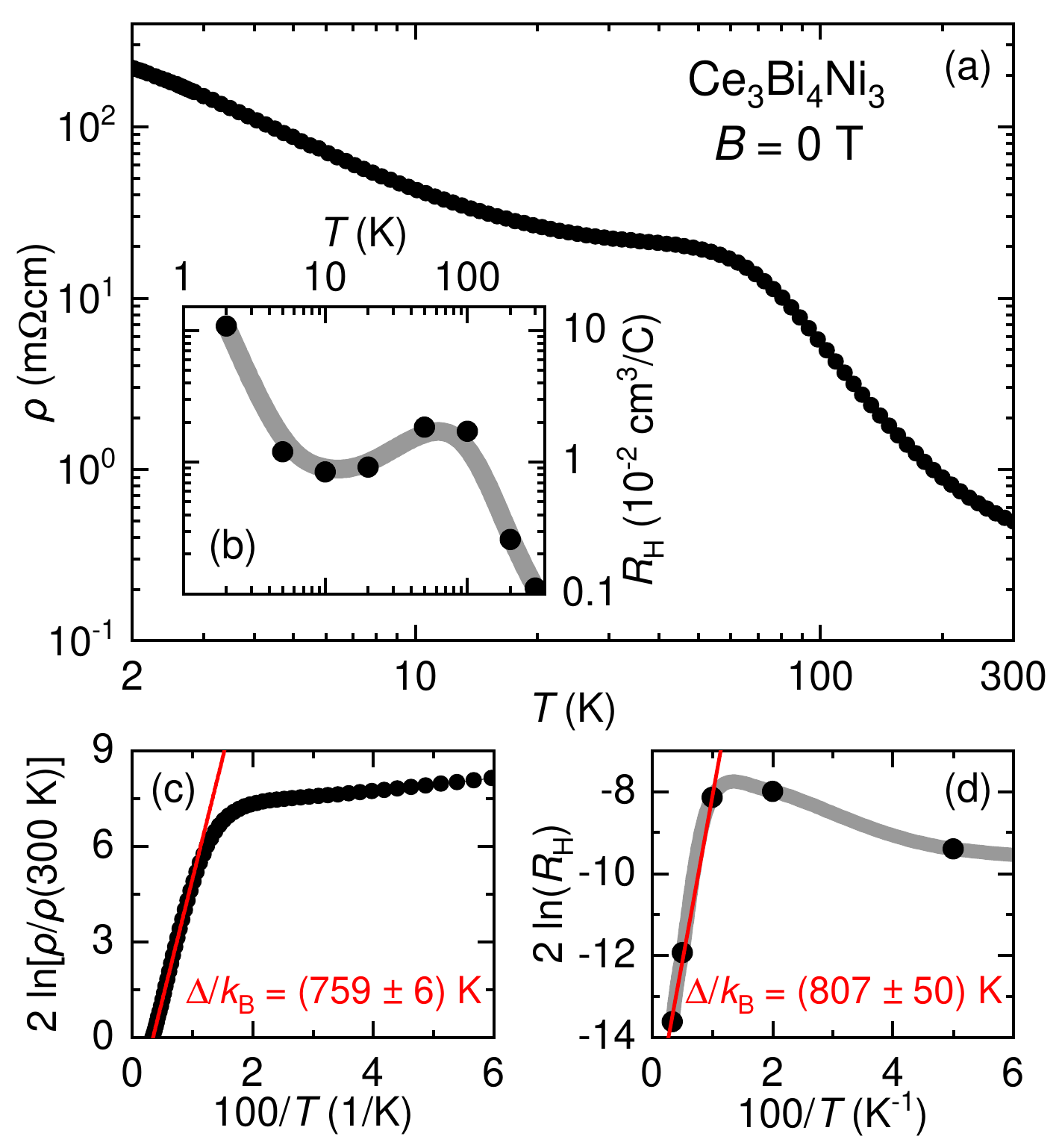}
\caption{\textbf{(a)} Electrical resistivity $\rho$ of \cebini\ as a function of temperature $T$ in zero magnetic field, with the electrical current applied along the crystallographic [111] direction. \textbf{(b)} Temperature-dependent Hall coefficient $R_\mathrm{H}$. The shaded grey line is a guide to the eye. \textbf{(c),(d)} Arrhenius plots of the resistivity (normalized to 300\,K) and Hall coefficient data. The red solid lines are linear fits to the data above 100\,K.}
\label{fig-transport}
\end{figure}

The electrical resistivity $\rho(T)$ of \cebini\ is displayed in Fig.\,\hyperref[fig-transport]{\ref{fig-transport}(a)}. Similarly to the Pt and Pd 3-4-3 compounds, nonmetallic behavior is observed across the entire studied temperature range (a detailed comparison is presented in Section \ref{discussion}). $\rho$ increases by more than two orders of magnitude from room temperature to 2\,K, with a resulting inverse residual resistivity ratio $iRRR$ = $\rho$(2\,K)/$\rho$(300\,K) = 440. Above 100\,K, $\rho(T)$ can be well described by a thermally activated scattering process across a gap. At lower temperatures, the system enters a regime where the increase in resistivity with decreasing temperature is less pronounced. First, a shoulder develops at approximately 50\,K, followed by a second, low-temperature thermally activated regime between 10\,K and 30\,K. A small energy gap of $\sim 20$\,K could be extracted from this narrow temperature range, although one might argue that other effects, such as the presence of in-gap states, could be responsible for this particular temperature dependence of the resistivity. Finally, $\rho(T)$ begins to flatten below 10\,K, but without reaching saturation down to 2\,K.

Figure \hyperref[fig-transport]{\ref{fig-transport}(b)} displays the temperature evolution of the Hall coefficient $R_\mathrm{H}$, obtained from linear fits to the low-field region of isothermal Hall resistivity curves (linear in $B \leq 9$\,T for 10\,K\,$\leq T$\,$\leq$\,300\,K). $R_\mathrm{H}$($T$) reaches a local maximum near 50\,K and tends to saturate at lower temperatures. This peak coincides with the end of the high-$T$ thermally activated regime. The increase of $R_\mathrm{H}(T)$ below 10\,K might be due to the freezing out of thermally activated carriers from shallow in-gap states, the origin of which needs further investigation.

To determine the size of the gap $\Delta$, the transport data are fitted with thermal activation functions, $\rho = \rho_0 \exp[\Delta/(2k_\mathrm{B}T)]$ and $R_{\rm H} = R_{\rm H,0} \exp[\Delta/(2k_\mathrm{B}T)]$, as shown in Arrhenius plots [Fig.\,\hyperref[fig-transport]{\ref{fig-transport}(c, d)}]. $\Delta$ values of 759\,K and 807\,K are obtained from the fits of the resistivity and Hall coefficient above 100\,K, respectively. These values are 7--10 times larger than the values reported for \cebipt\ \cite{hundley90a, dzsaber17a} and at least 45 times larger than the Kondo gap acting on the background of the WKSM state of \cebipd\ in the absence of applied magnetic field \cite{dzsaber17a,dzsaber22a}, suggesting the presence of a much larger Kondo energy scale at play in the Ni compound. 

\begin{figure}[t]
\centering
\includegraphics[width=0.6\columnwidth]{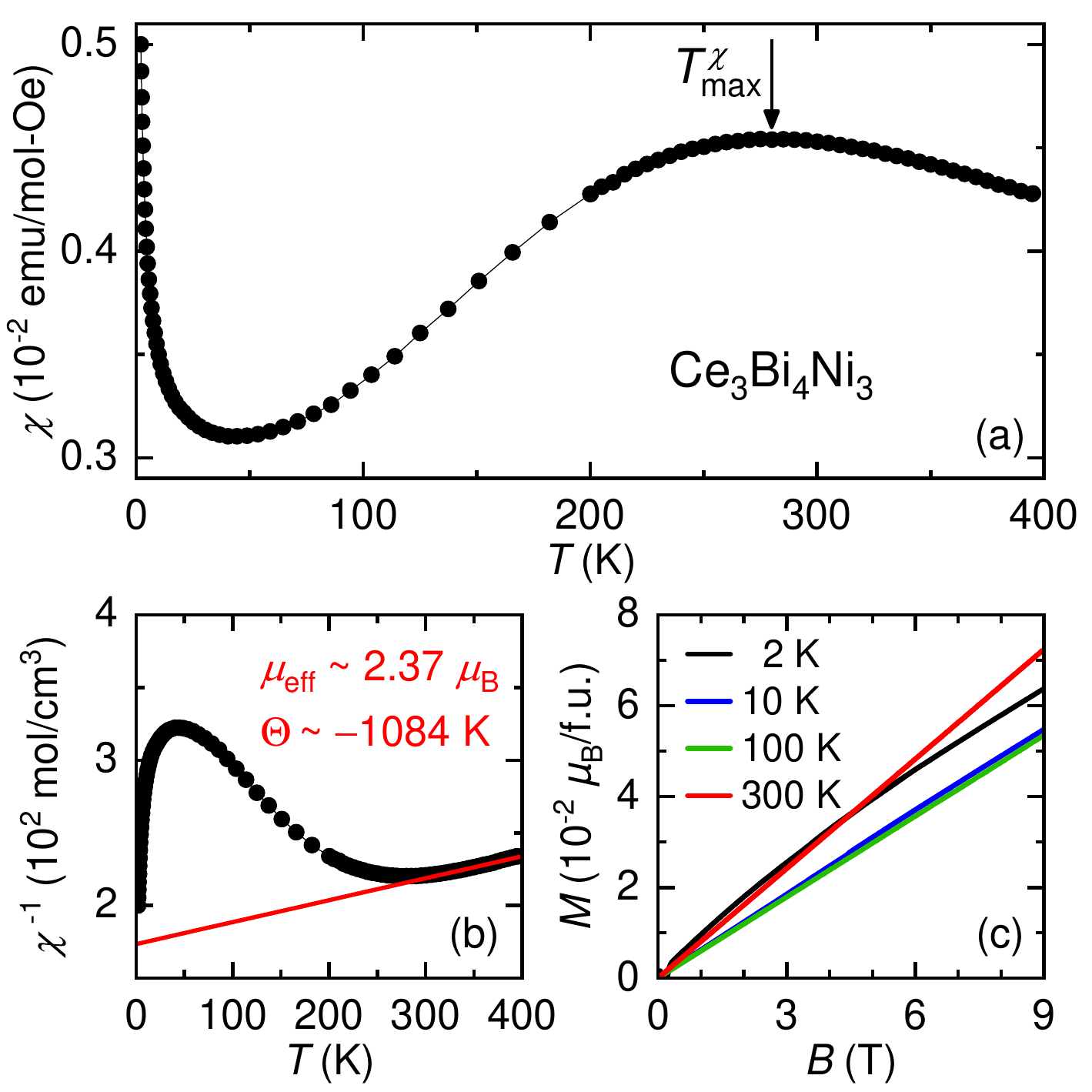}
\caption{\textbf{(a)} Temperature dependence of the magnetic susceptibility $\chi$ of \cebini, measured with $B = 1$\,T. \textbf{(b)} Inverse magnetic susceptibility $\chi^{-1}(T)$. The paramagnetic Weiss temperate $\Theta$ and the effective magnetic moment $\mu_\mathrm{eff}$ were determined from a Curie-Weiss fit for $T > T^{\chi}_{\mathrm{max}}$ (red line). \textbf{(c)} Magnetization $M$ vs $B$ at 2\,K (black), 10\,K (blue), 100\,K (green) and 300\,K (red). The curves show up and down $B$-sweeps, with no signs of magnetic hysteresis detected. For all cases, $B$ was applied parallel to the crystallographic [11-2] direction.}
\label{fig-chi}
\end{figure}
The magnetic susceptibility $\chi(T)$ of \cebini\ displays a broad maximum at $T^{\chi}_{\mathrm{max}} = 280$\,K [Fig.\,\hyperref[fig-chi]{\ref{fig-chi}(a)}]. In Kondo insulators, the occurrence of such maxima has been associated with the onset of Kondo screening and the opening of a hybridization gap \cite{riseborough00a}. We estimate a Kondo temperature using the Kondo lattice expression $T_\mathrm{K} \approx 4T^{\chi}_\mathrm{max}/(2J + 1) = 187$\,K \cite{coleman83a, hundley90a}, with a total angular momentum $J = 5/2$ for Ce$^{3+}$. The upturn in $\chi$ vs $T$ observed at lower temperatures is generally attributed to small amounts of magnetic impurities or defects \cite{takabatake98a} which were not detected in our XRD experiments. Assuming Curie-Weiss behavior for $T > T^{\chi}_{\mathrm{max}}$, a linear fit to the inverse susceptibility $\chi^{-1}(T)$ in the limited available temperature range [Fig.\,\hyperref[fig-chi]{\ref{fig-chi}(b)}] yields a large negative Weiss temperature $\Theta \sim -1084$\,K ---typical for Kondo insulators--- and an effective magnetic moment $\mu_{\mathrm{eff}} \sim 2.37 \mu_{\mathrm{B}}$, slightly reduced from the Hund's rules value of $2.54 \mu_{\mathrm{B}}$ for free Ce$^{3+}$ ions.

The magnetization curves $M(B)$ are linear in field at 10\,K, 100\,K, and 300\,K [Fig.\,\hyperref[fig-chi]{\ref{fig-chi}(c)}]. A slight curvature appears only in the 2\,K measurement, without reaching saturation up to 9\,T. No indications of magnetic hysteresis were observed in any of the measurements. Overall, \cebini\ appears to be well described as a paramagnetic Kondo insulator, without magnetic ordering down to 2\,K.

\begin{figure}[t]
\centering
\includegraphics[width=0.6\columnwidth]{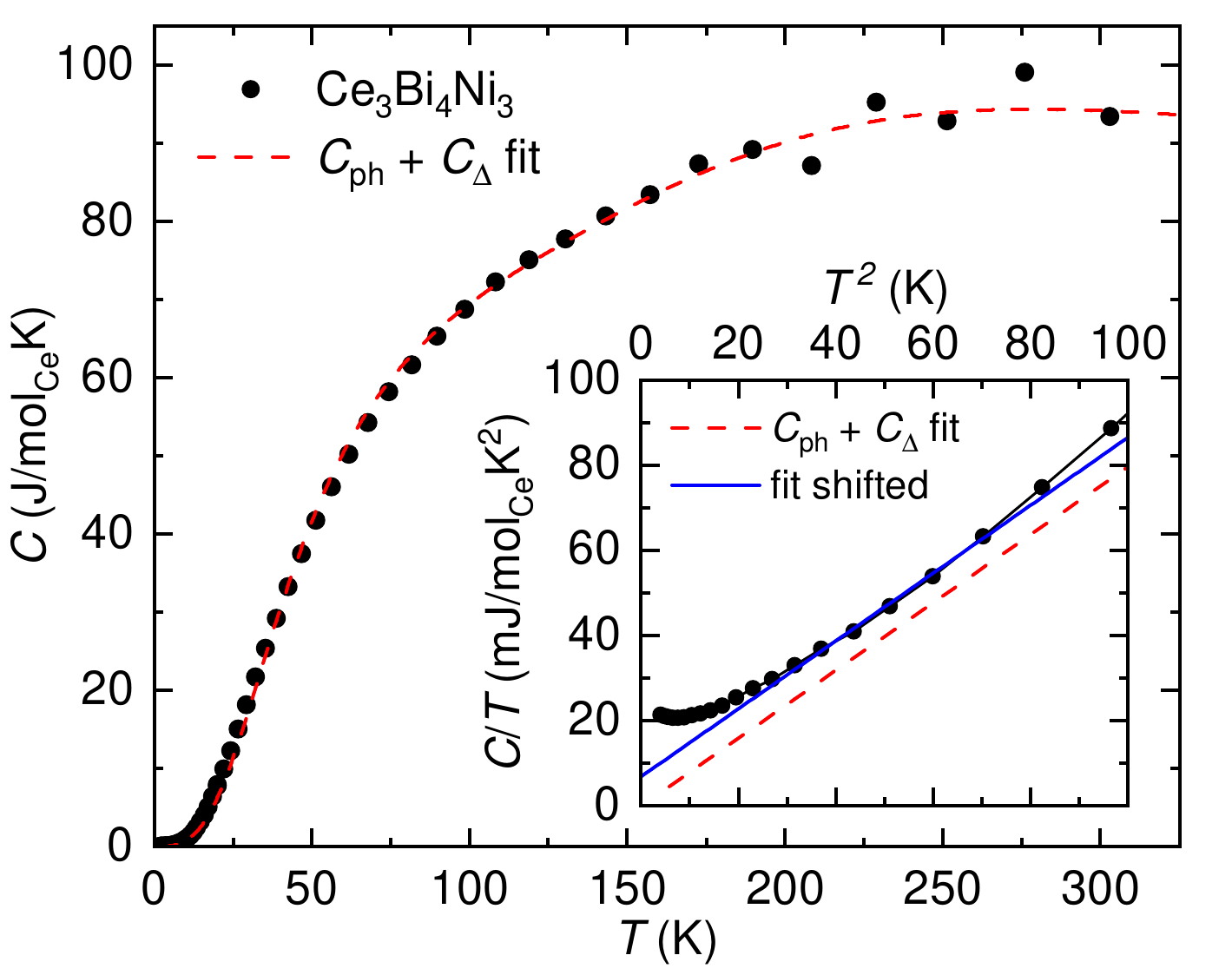}
\caption{Specific heat $C$ vs $T$ of \cebini\ measured between 2\,K and 300\,K. The red dashed line is a fit with the Debye model plus an additional Schottky-like electronic contribution derived for Kondo insulators (see main text). The inset shows the low temperature $C/T$ vs $T^2$ and the fit. The Sommerfeld coefficient ($\gamma$) is then determined by shifting the fitting line by $\gamma = 6.92$\,mJ/mol$_\mathrm{Ce}$K$^2$ to match the measured data (blue line).}
\label{fig-Cp}
\end{figure}

We can gain further understanding on the nature of the ground state of \cebini\ and the strength of the Kondo interaction from the specific heat $C(T)$ measurements presented in Fig.\,\ref{fig-Cp}. So far, our efforts to synthesize La$_3$Bi$_4$Ni$_3$ as a non-$f$-electron reference compound have been unsuccessful. To isolate the electronic and magnetic (Ce 4$f$ related) contributions to the specific heat, we therefore model the data as $C(T) = \gamma T + C_{\rm ph} + C_{\Delta}$. The first term is the linear-in-temperature electronic (Sommerfeld) contribution, the second term the phonon specific heat described by the Debye model \cite{kittel04a}. $C_\Delta$ accounts for an additional Schottky-like electronic contribution in Kondo insulators, which has previously been modelled with two sharp peaks in the density of states, located above and below the energy gap \cite{riseborough00a}:
\begin{equation}
C_\Delta \propto \left(\frac{\Delta}{2k_\mathrm{B}T}\right)^2 \frac{(2J+1) \exp\left[\Delta/2k_\mathrm{B} T\right]}{\left(2J+1 + \exp\left[\Delta/2k_\mathrm{B} T\right]\right)^2}\; .
\end{equation}

The contribution from the Sommerfeld term is expected to be very small.  In addition, its extraction is complicated by a low-temperature upturn in $C/T$ (see inset of Fig.\,\ref{fig-Cp}). We therefore set $\gamma = 0$ and fit the data above 20\,K using the full Debye integral. The fit yields a Debye temperature $\Theta_{\rm D} = (202 \pm 3)$\,K and a large gap $\Delta = (1693 \pm 87)$\,K. We can nevertheless obtain an estimate of the Sommerfeld coefficient $\gamma$ by matching the fit to the measured data in a $C/T$ vs $T^2$ plot at low temperatures (blue line in inset of Fig.\,\ref{fig-Cp}). The required shift along the $C/T$ axis then corresponds to $\gamma = 6.92$\,mJ/mol$_\mathrm{Ce}$K$^2$, similarly small to the value determined for \cebipt\ \cite{hundley90a}. The small low-temperature upturn visible in $C/T$ could be due to magnetic impurities, as we previously discussed for the magnetic susceptibility data (Fig.\,\ref{fig-chi}). However, the upturn could also be related to the onset of non-Fermi liquid behavior in the presence of a nearby quantum critical point. Additional low-temperature investigations are needed to unveil its true origin.

\section{Discussion}\label{discussion}

We can now take a look at \cebini\ in the context of the \cebix\ series ($X$ = Ni, Pd, Pt). The lattice parameters $a$ for the three compounds are shown in Fig.\,\hyperref[fig-powderXRD]{\ref{fig-powderXRD}(e)}. In addition, we include the values corresponding to the intermediate Pt-Pd substitution Ce$_{3}$Bi$_{4}$(Pt$_{1-x}$Pd$_x$)$_{3}$ from Ref.\,\cite{dzsaber17a} and one value corresponding to the intermediate Pd-Ni substitution Ce$_{3}$Bi$_{4}$(Pd$_{1-x}$Ni$_x$)$_{3}$ with $x = 0.19$. Unlike the Pt-Pd case, which appears to exert very little to no-effect on the unit cell volume \cite{dzsaber17a}, the substitution of Pd with Ni is no longer isosize. In this context, chemical (and, possibly, external) pressure becomes now a relevant parameter for tuning the Kondo coupling $J_\mathrm{K}$ across the extended Ce 3-4-3 series.

A recent study on the electronic transport properties of \cebipd\ under hydrostatic pressure showed that the energy gap extracted from resistivity increases quadratically with applied pressure, reaching a value of $\sim 4.8$\,meV at 2.3\,GPa \cite{ajeesh22a,note_ajeesh}. Hence, pressure appears to drive \cebipd\ to a more insulating state. In this work \cite{ajeesh22a}, as well as in two recent theoretical studies \cite{xu22a, zhang22a}, it was thus argued that pressure drives \cebipd\ towards \cebipt, which, at that point, was the only known isoelectronic compound of this series. In view of the chemical pressure effect presented in Fig.\,\hyperref[fig-powderXRD]{\ref{fig-powderXRD}(e)}, here we argue that, instead, pressure drives \cebipd\ towards the Ni compound. We can translate the 8.3\% shrinking of the unit cell volume obtained from \cebipd\ to \cebini\ to an equivalent externally applied pressure $P$ using the third-order Birch-Murnaghan equation of state \cite{birch47a}
\begin{equation}
P(V)=\frac{3B_0}{2} \left[ \left( \frac{V_0}{V} \right)^{\frac{7}{3}} - \left( \frac{V_0}{V} \right)^{\frac{5}{3}} \right] \times \left\{ 1 + \frac{3}{4}(B'_0 -4) \left[ \left( \frac{V_0}{V} \right)^{\frac{2}{3}} - 1 \right] \right\},
\end{equation}
where $V_0$ is the ambient pressure volume, $V$ is the compressed volume, $B_0$ is the bulk modulus, and $B'_0$ its derivative with respect to pressure. $B_0$ and $B'_0$ result typically from fits to $V(P)$ data obtained in high-pressure XRD experiments. Since such data are not yet available for \cebipd, we use the values $B_0 = 59.1$\,GPa and $B'_0 = 6.9$ reported for the isostructural (and isosize) compound \cebipt\ \cite{campbell19a}, which provide also a good agreement with the pressure-dependent unit cell parameters calculated via DFT+U \cite{zhang22a}. This calculation yields a pressure of 6.9\,GPa equivalent to the compression effect resulting from the Pd-Ni substitution. Using the quadratic dependence of $\Delta(P)$ from the experiments reported in Refs.\,\cite{ajeesh22a,note_ajeesh}, a pressure of 9.4\,GPa would then be necessary to reach the energy gap value estimated for \cebini\ (Fig.\,\ref{fig-transport}), close to the value calculated using the Birch-Murnaghan formula.
\begin{figure}[t]
\centering
\includegraphics[width=0.6\columnwidth]{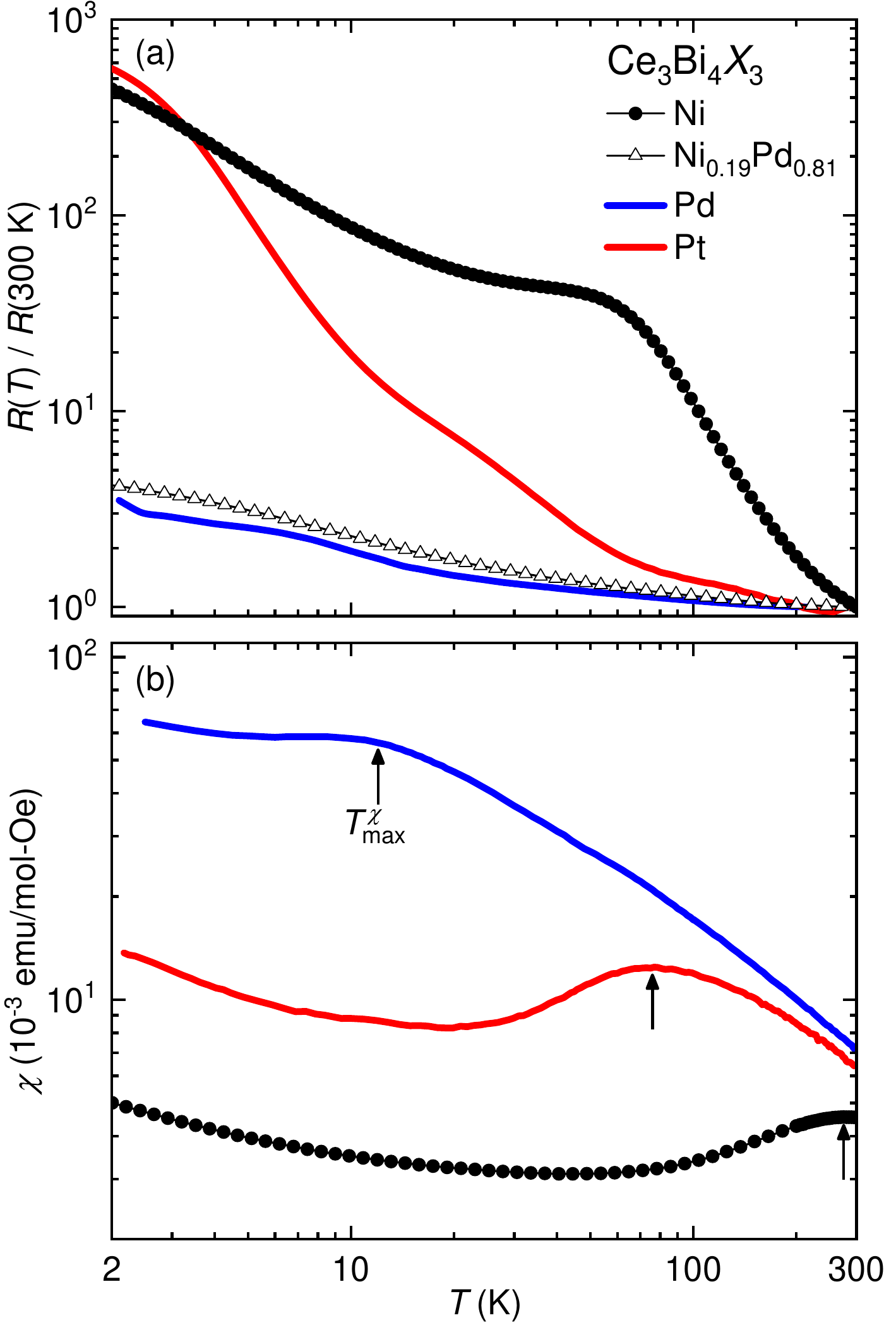}
\caption{\textbf{(a)} Electrical resistance $R(T)$ normalized to its room temperature value $R$(300\,K) and \textbf{(b)} magnetic susceptibility $\chi(T)$ of the \cebix\ series, with $X$ = Ni (black dots), Pd (blue line), Pt (red line), and of a partially substituted Ni-Pd single crystal (19\% Ni, white triangles). The data for \cebipd\ and \cebipt\ were taken from Refs.\,\cite{dzsaber17a} and \cite{dzsaber21a}.}
\label{fig-343X}
\end{figure}

\begin{figure}[t]
\centering
\includegraphics[width=0.58\columnwidth]{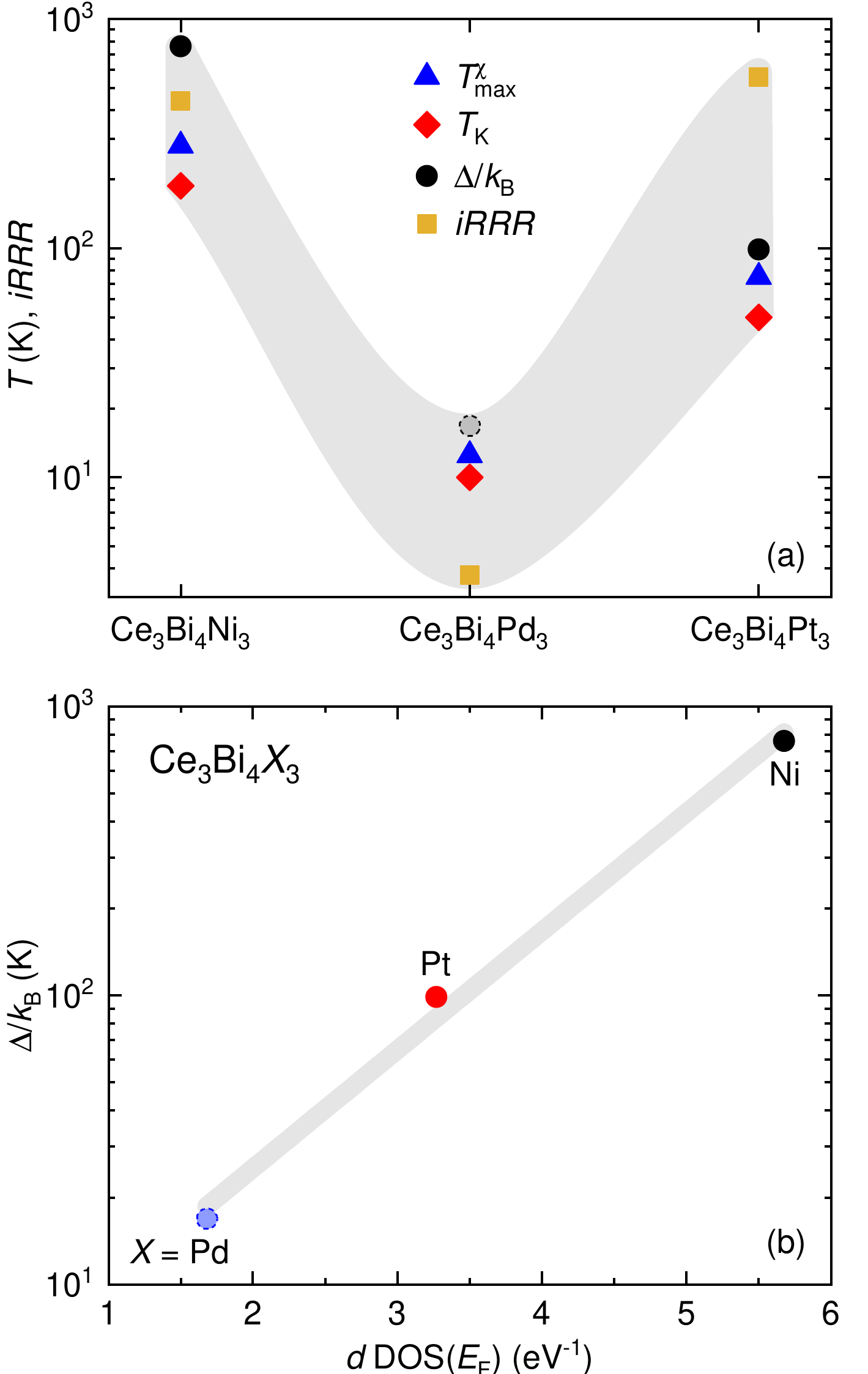}
\caption{Comparison of characteristic parameters for the \cebix\ series ($X= $\ Ni, Pd, Pt). Shaded grey areas are guides to the eyes. \textbf{(a)} $iRRR$ values (from Fig.\,\ref{fig-343X}) and temperature scales $T_\mathrm{K}$, $\Delta/k_\mathrm{B}$, and $T^{\chi}_\mathrm{max}$. For \cebini, $T^{\chi}_\mathrm{max}$ is taken from Fig.\,\hyperref[fig-chi]{\ref{fig-chi}(a)}, $T_\mathrm{K}$ is calculated from $T^{\chi}_\mathrm{max}$ (see Section \ref{results}), and $\Delta/k_\mathrm{B}$ is obtained from the fits in Figs.\,\hyperref[fig-transport]{\ref{fig-transport}(c-d)}. Values for the Pd and Pt compounds taken from Ref.\,\cite{dzsaber17a}. \textbf{(b)} Kondo energy gap $\Delta/k_\mathrm{B}$ as a function of the $d$-electron density of states ($d$\,DOS) at the Fermi level (from DFT calculations, see Figs.\,\ref{fig-DFT} and \ref{fig-DOS} in Appendix \ref{apx-DOS}).}
\label{fig-343Xscales}
\end{figure}

In Fig.\,\ref{fig-343X} we compare the normalized resistances $R(T)/R(\mathrm{300\,K})$ and magnetic susceptibilities of the Ni, Pd, and Pt compounds. Perhaps surprisingly at first sight, the transport and magnetic properties of this isoelectronic series evolve nonmonotonically with atomic number. The $iRRR$ decreases by more than two orders of magnitude from \cebini\ to \cebipd, and recovers the value of the Ni compound when Pd is substituted by Pt. A slight change in resistivity from the pure Pd compound is already evident when Pd is substituted by 19\% Ni [blue line vs white triangles in Fig.\,\hyperref[fig-343X]{\ref{fig-343X}(a)}]. Similarly, the characteristic peak in susceptibility at $T^{\chi}_{\mathrm{max}}$ [indicated by arrows in Fig.\,\hyperref[fig-343X]{\ref{fig-343X}(b)}] shifts first from room temperature to $\sim 12$\,K from the Ni to the Pd compound, and increases back to $\sim 75$\,K for \cebipt. 

To help visualizing the evolution across the series, we present in Fig.\,\hyperref[fig-343Xscales]{\ref{fig-343Xscales}(a)} (notice the logarithmic vertical axis) the values of $T^{\chi}_{\mathrm{max}}$ and $iRRR$ extracted from Fig.\,\ref{fig-343X}, together with the values $\Delta/k_\mathrm{B}$ and $T_{\mathrm{K}}$ obtained as described in Section \ref{results}. As we mentioned earlier, the transport properties of \cebipd\ at low temperatures and zero magnetic field are largely dominated by the effects of the gapless WKSM ground state \cite{dzsaber17a,dzsaber21a,dzsaber22a}. For this reason, as noted in Ref.\,\cite{dzsaber17a}, the finite energy gap value of 16.9\,K for \cebipd\ obtained via Arrhenius fits to the $B = 0$ resistivity data should be taken with caution, and therefore it is presented as a shaded/dotted line symbol in Fig.\,\ref{fig-343Xscales}. Nevertheless, the value is in good agreement with the magnetic field-derived energy of $\sim 1.5$\,meV required to close the Kondo gap \cite{dzsaber22a}.

Overall, the nonmonotonic change of the characteristic parameters forms a ``U'' shape centered around the most strongly correlated material in the series \cebipd, and with the comparably less correlated, large Kondo-gap compounds \cebipt\ and \cebini\ lying at either side. While pressure arguments alone could suffice, as discussed above, to broadly describe the Ni-Pd side of this plot, they cannot explain the changes in Kondo coupling observed under the volume-preserving Pd-Pt substitution, and, therefore,  other arguments are necessary to understand the entire nonmonotonic evolution found across the series. A recent analysis of electronic band structure calculations suggested that the radial extension of the $d$ orbitals could play an important role in the determination of the Kondo coupling in these hybridization-gapped materials \cite{tomczak20a}. Our density functional theory (DFT) calculations of the whole series (Fig.\,\ref{fig-DFT}) appear to confirm this hypothesis. The aim of these calculations is to extract the $d$-electron density of states at the Fermi level, a key factor determining the Kondo interaction strength. As the Kondo effect itself cannot be captured by DFT, the $4f$ electron of Ce is treated as core electron (see Appendix \ref{apx-DOS} for more details). Figure \hyperref[fig-343Xscales]{\ref{fig-343Xscales}(b)} displays the Kondo scale $\Delta/k_\mathrm{B}$ as a function of the $d$-orbital electron density of states ($d$\,DOS) at the Fermi level for the three \cebix\ compounds (see Fig.\,\ref{fig-DOS} in Appendix \ref{apx-DOS}). The Kondo scale increases exponentially with the $d$\,DOS($E_{\mathrm{F}}$) (shaded grey line in the semi-log plot), but it does it nonmonotonically with the principal atomic number $n$, increasing from the Pd 4$d$ to the Pt 5$d$ compounds, and reaching its maximum for the 3$d$ \cebini. Hence, we conclude that the nonmonotonic evolution of the physical properties across the isoelectronic series can be explained by the relationship between Kondo coupling and $d$-orbital density of states.

\section{Conclusions}\label{conclusion}
In summary, we successfully synthesized single crystals of \cebini\ using Pb flux. The high quality of the samples was verified via XRD and EDX measurements. From electrical resistivity, Hall effect, magnetic susceptibility, and specific heat experiments, we conclude that \cebini\ is a paramagnetic heavy fermion material with a relatively large Kondo gap of 65--70\,meV. In the context of the \cebix\ series ($X$ = Ni, Pd, Pt) we find a positive chemical pressure effect in \cebini\ relative to the heavier Pd and Pt compounds. Notably, the transport and magnetic properties as well as the extracted characteristic energy scales evolve nonmonotonically throughout this isoelectronic series. We attribute this behavior to the nonmonotonic variation of $d$-orbital DOS at the Fermi level. Obtaining a detailed understanding of how the variations in orbital hybridization affect the Kondo coupling across the Ni-Pd-Pt isoelectronic series would have a broader impact for future experimental and theoretical studies. For example, the giant topological responses found in various physical properties of \cebipd\ arise from strongly renormalized Weyl bands located within a Kondo insulating gap \cite{dzsaber17a,lai18a,dzsaber21a}, which in turn acts in the background as a shield against the effects of topologically trivial bands. It would then be interesting to explore how the WKSM signatures observed in \cebipd\ will evolve as the band structure is carefully tuned via various Ni-Pd-Pt substitutions, and as a function of externally applied pressure.

\begin{acknowledgments}
The work in Vienna was supported by the Austrian Science Fund (FWF projects I\,4047-N27 and I\,5868-FOR 5249 - QUAST), the European Microkelvin Platform (H2020 project No.\ 824109), and the European Research Council (ERC Advanced Grant No.\ 101055088, CorMeTop). The work in Prague was supported by the Czech Science Fund (GA\v{C}R project No.\ 19-29614L). M.G.V.\ acknowledges support to the Spanish Ministerio de Ciencia e Innovaci\'{o}n (Grant No.\ PID2022-142008NB-I00), partial support from European Research Council (ERC) Grant Agreement No.\ 101020833 and the European Union NextGenerationEU/PRTR-C17.I1, as well as by the IKUR Strategy under the collaboration agreement between Ikerbasque Foundation and DIPC on behalf of the Department of Education of the Basque Government. D.-G.O., C.F., R.V., and M.G.V.\ acknowledge funding from the German Research Foundation (DFG) and the Austrian Science Fund through the research unit FOR 5249 - QUAST.
\end{acknowledgments}

\clearpage

\appendix
\section{Materials and Methods}\label{apx-methods}

Single crystal and polycrystalline samples of \cebini\ were synthesized using high purity ingots of Ce (Smart Elements, 99.9\%), Bi (Aldrich, 99.999\%), and Ni (Strem Chemicals, 99.9\%). As a first step, a polycrystalline sample with a nominal composition of \cebini\ was prepared in a high-frequency furnace. From XRD and SEM/EDX measurements (not presented here), the as-cast sample contained the desired \cebini\ phase (Y$_{3}$Sb$_{4}$Au$_{3}$-type, space group (SG) 220, $I\overline{4}$3$d$), additional CeBi (NaCl-type, SG:\,225) and CeNi$_{1-x}$Bi$_{2}$ (CuHfSi$_{2}$-type, SG:\,129) phases, and small amounts of elemental Ni. After annealing at 700\,$^{\circ}$C in vacuum for 14 days, CeNi$_{1-x}$Bi$_{2}$ disappeared in the ingot, but the binary phase remained. From a careful examination of the microstructure of the as-cast sample, we concluded that the cubic CeBi phase crystallizes before the \cebini\ phase during the cooling process, i.e.\ \cebini\ melts incongruently.

For the single crystal growth various metal fluxes were tested, including Al, Ga, In, Sn, Sb, Bi, and Pb. Only with the latter (Pb, Alfa Aesar 99.999\%) single crystals of \cebini\ successfully formed. The high-purity elements were loaded into a tantalum crucible with a ratio of Ce:Bi:Ni:Pb = 3:4:3.2:10 and sealed in vacuum in a quartz ampoule. The excess of Ni was used to suppress its deficiency in the 3-4-3 phase (see EDX analysis in Section \ref{crystalgrowth} and discussion in Appendix \ref{apx-stoichiometry}). The ampoule was heated in a vertical furnace to 1050\,$^{\circ}$C at 150\,$^{\circ}$C/h, and held during 10\,h for soaking before cooling it down to 500\,$^{\circ}$C at 3\,$^{\circ}$C/h. The ampoule was then inverted and centrifuged to separate the single crystals from the flux. In addition to the desired \cebini\ phase, we also detected the phases CeBi and CeNi$_{1-x}$Bi$_{2}$. While the former appears to be unavoidable, it can be safely separated from the selected \cebini\ single crystals. The formation of the latter, on the other hand, could be circumvented by removing the crucible from the furnace at a certain high temperature.

The phase constitutions and lattice parameters of all specimens were analyzed by powder XRD using Ge-monochromated Cu-K$\upalpha_{1}$ radiation ($\lambda = 1.5406$\,\AA). Rietveld refinements were obtained with the FullProf software suite \cite{rodriguez93a,roisnel01a} and precise lattice parameters with the program Struktur \cite{wacha89a} from least-squares fits; Ge served as an internal standard ($a_{\mathrm{Ge}} = 5.657906$\,\AA). The refinement described in Section \ref{crystalgrowth} was achieved with reasonable residual values, R$_{\mathrm{B}} = 7.4$\%, R$_{\mathrm{F}} = 7.6$\%, and a goodness of fit of 1.6.

Small single crystals ($\sim 50\,\upmu$m) suited for x-ray structure analysis were selected from a mechanically crashed large crystal. The small crystals were firstly inspected on an AXS D8-GADDS texture goniometer, to assure the high crystal quality, unit cell dimensions and Laue symmetry. Detailed single crystal XRD data were collected at 100\,K, 200\,K, and 300\,K on a Bruker KAPPA Apex II four-circle diffractometer equipped with CCD detector (graphite monochromated Mo-K$_{\upalpha}$ radiation, $\lambda = 0.7107$\,\AA). Besides the general treatment of absorption effects using the multi-scan technique (SADABS; redundancy of integrated reflections $> 8$) \cite{bruker04a}, no individual absorption correction was necessary because of the rather regular crystal shape and small dimensions of the investigated specimens. The crystal structure was solved by applying direct methods (software SHELXS-97) and refined against F$^2$ (software SHELXL-97-2) within the Oscail program \cite{farrugia99a,mcardle04a,sheldrick08a}. Finally, the crystal structure was standardized with the Structure Tidy software \cite{gelato87a,parthe93a}. Details on structure refinements of the studied phases are listed in Appendix \ref{apx-structure}. Crystal orientations were determined and the quality of the samples was verified by Laue XRD (Photonic Science).

The chemical composition was determined by EDX in a scanning electron microscope operated at 20\,kV (Zeiss Supra 55VP, probe size 1\,$\upmu$m). Measurements at different positions on the polished surface of the samples yielded a variation of composition homogeneity of $\pm 0.3$\,at.\,\%.

Electrical transport and magnetic susceptibility measurements were performed between 300\,K and 2\,K in a PPMS/VSM system from Quantum Design. Electrical contacts for resistivity were made along the [111] crystallographic direction by gluing 25\,$\upmu$m annealed gold wires with silver paste (DuPont 4929N). The heat capacity was also measured in a PPMS system using the standard heat-pulse method. Samples were fixed to the platform with Apiezon N-grease. Before the measurements of the samples, the empty platform was calibrated with N-grease (addenda).

\section{Structure refinement from powder and single crystal x-ray diffraction}\label{apx-structure}

\begin{figure}[h]
\centering
\includegraphics[width=0.5\columnwidth]{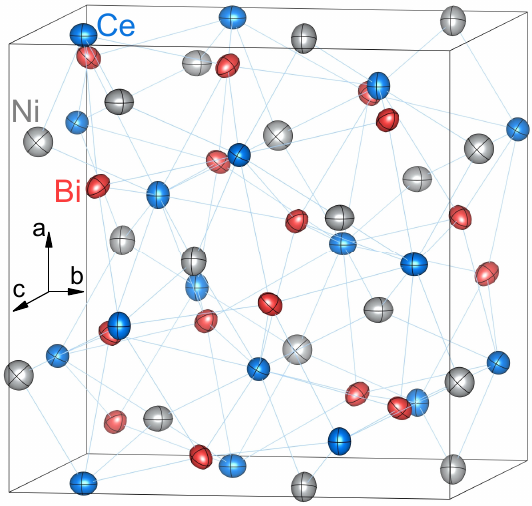}
\caption{Crystal structure of \cebini, with atoms displayed as ellipsoids representing their anisotropic atomic displacement parameters (ADPs). Derived from the refinement (including Ni deficiency in the structure) of single crystal XRD data corresponding to a highly stoichiometric sample (see Table \ref{table2}).}
\label{fig-ADPs}
\end{figure}

The Rietveld refinement of the powder XRD shows that the atomic displacement parameter (ADP) at the Ni site (12$b$) is larger than those at the corresponding Ce (12$a$) and Bi (16$c$) sites. This could originate from an inappropriate occupation at the Ni site. To clarify it, we carried out single crystal XRD at 100\,K, 200\,K, and 300\,K, expecting a priori that the influence of ADPs on the determination of occupations would be reduced at low temperatures. The single crystal XRD analysis reveals isomorphism of the Y$_{3}$Sb$_{4}$Au$_{3}$-type structure (SG:\,220, $I\overline{4}$3$d$), with Ce sitting at the 12$a$ site, Ni at the 12$b$ site, and Bi at the 16$c$ site. For samples corresponding to the most stoichiometric batch (\texttt{\#}36, see Appendix \ref{apx-stoichiometry}), a refinement of the atomic occupations at each site without any constraints yields almost full occupations at the Ce and Bi sites, but an occupation that slightly deviates from 1 at the Ni site (occupation $\sim 0.95$). This result is robust and does not depend on temperature (see Table\,\ref{table2}). The refinement including Ni deficiency in the structure (Table\,\ref{table1}) also yields reasonable ADPs at each site, low residual values, low residual electron densities, and a refined composition similar to that obtained from the EDX experiments [Ce$_{3.15(5)}$Bi$_{3.90(9)}$Ni$_{2.94(4)}$, see Section \ref{crystalgrowth}]. In the crystal structure displayed in Fig.\,\ref{fig-ADPs}, the shapes of the atoms represent the anisotropic ADPs, with almost spherical shapes and equal size for each atom. Table\,\ref{table3} lists the interatomic distances between all atoms, showing strong bonding between heterogeneous atoms if we compare the interatomic distances with atom radii of $r_{\mathrm{Ce}} = 1.83$\,\AA, $r_{\mathrm{Bi}} = 1.60$\,\AA, and $r_{\mathrm{Ni}} = 1.25$\,\AA.
 
Using this model in the refinement of the powder XRD data (red line in Fig.\,\ref{fig-powderXRD}), we obtained a better fit for the observed pattern, more reasonable ADPs at each site, and slightly lower residual values. To further establish the level of Ni deficiency in the structure, additional structure investigations with crystals of various compositions, which can be realized by modifying the crystal growth parameters, are necessary. 

\begin{table*}[h]
\centering
\caption{Structural data of \cebini\ obtained from single crystal XRD data collected at 100\,K, 200\,K, and 300\,K (space group 220, $I\overline{4}$3$d$). For the refinement, the lattice parameter at 300\,K was initially obtained from powder XRD, and the values at lower temperatures were calculated from the linear thermal expansion coefficient derived from single crystal  XRD.}
\vspace{0.2cm}
\begin{threeparttable}
\begin{tabular}{l c c c}
\hline \hline 
$T$ (K) &  100\,K & 200\,K & 300\,K \\		
\hline
$a$ (\AA) & 9.7170(2) & 9.7445(2) & 9.7720(2) \\
$V$ (\AA$^3$) & 917.48 & 925.29 & 933.15 \\
Density (g cm$^{-3}$) & 10.370 & 10.282 & 10.196 \\
$\mu_{\mathrm{abs}}$ (mm$^{-1}$) & 96.88 & 96.06 & 95.25 \\				
$F(000)$ & 2360.0 & 2360.0 & 2360.0 \\
Reflections & 14509 & 14555 & 14993 \\
Independent reflections & 657 & 657 & 672 \\  
$2\theta_\mathrm{max}$ & 90.63 & 90.30 & 91.19 \\ 
$R_\mathrm{int}$ & 0.069 & 0.073 & 0.073 \\ 
$h$ & $-19 \leq h \leq 19$ & $-19 \leq h \leq 19$ & $-19 \leq h \leq 19$ \\ 
$k$ & $-19 \leq k \leq 11$ & $-19 \leq k \leq 11$ & $-19 \leq k \leq 11$ \\ 
$l$ & $-19 \leq l \leq 17$ & $-19 \leq l \leq 17$ & $-19 \leq l \leq 17$ \\ 
Reflections in refinement & $630(F_o) > 4\upsigma(F_o)$ & $623(F_o) > 4\upsigma(F_o)$ & $630(F_o) > 4\upsigma(F_o)$ \\ 
Number of variables & 10 & 10 & 10 \\ 
Extinction coefficient & 0.0023(2) & 0.0036(2) & 0.0039(3) \\ 
Goodness of fit & 1.25 & 1.22 & 1.21 \\ 
$R_\mathbf{1}$ & 0.021 & 0.019 & 0.020 \\ 
$wR_2$ & 0.053 & 0.058 & 0.062 \\ 
$\Delta\rho_\mathrm{max}, \Delta\rho_\mathrm{min}$ (e\AA$^{-3}$) & 3.23, $-3.88$ & 2.68, $-3.78$ & 2.48, $-3.70$ \\ 
Refined composition & Ce$_{3}$Bi$_{4}$Ni$_{2.85(3)}$ & Ce$_{3}$Bi$_{4}$Ni$_{2.85(3)}$ & Ce$_{3}$Bi$_{4}$Ni$_{2.82(6)}$ \\				
\hline \hline
\end{tabular}
\end{threeparttable}
\label{table1}	
\end{table*}

\begin{table*}
\centering
\caption{Structural information and atomic displacement parameters $U_{ij}$ in [$10^2$\,nm$^2$] at 100\,K, 200\,K, and 300\,K.}
\vspace{0.2cm}
\begin{tabular}{l c c c}	
\hline \hline
Wyckoff site & 100\,K &  200\,K &  300\,K \\ 
\hline
$12a$ (\nicefrac{3}{8}, 0, \nicefrac{1}{4}): \textit{Occ.} & Ce:\,1 & Ce:\,1 & Ce:\,1 \\  
\hspace{5mm} $U_{eq}$ (\AA$^2$) & 0.0009(1) & 0.0033(1) & 0.0063(1) \\ 
\hspace{5mm} $U_{11}$ (\AA$^2$) & 0.0015(2) & 0.0045(2) & 0.0079(3) \\ 
\hspace{5mm} $U_{ii}$ (\AA$^2$), $i = 2, 3$ & 0.0006(1) & 0.0028(2) & 0.0055(2) \\ 
\midrule
$12b$ (\nicefrac{7}{8}, 0, \nicefrac{1}{4}): \textit{Occ.} & Ni:\,0.95(1) & Ni:\,0.95(1) & Ni:\,0.95(1) \\  
\hspace{5mm} $U_{eq}$ (\AA$^2$) & 0.0019(4) & 0.0049(5) & 0.0083(5) \\ 
\hspace{5mm} $U_{11}$ (\AA$^2$) & 0.0010(6) & 0.0038(7) & 0.0066(8) \\ 
\hspace{5mm} $U_{ii}$ (\AA$^2$), $i = 2, 3$ & 0.0023(5) & 0.0054(5) & 0.0092(6) \\ 
\midrule
$16c$ ($x, x, x$): \textit{Occ.} & Bi, $x = 0.07929(2):1$ & Bi, $x = 0.07893(2):1$ & Bi, $x = 0.07929(2):1$ \\  
\hspace{5mm} $U_{eq}$ (\AA$^2$) & 0.0007(1) & 0.0031(1) & 0.0059(1) \\ 
\hspace{5mm} $U_{ii}$ (\AA$^2$), $i = 1, 2, 3$ & 0.0007(1) & 0.0031(1) & 0.0059(1) \\ 
\hline \hline
\end{tabular}
\label{table2}	
\end{table*}

\begin{table*}[h]
\centering
\caption{Interatomic distances (\AA) at 100\,K, 200\,K, and 300\,K.}
\vspace{0.2cm}
\begin{tabular}{c c c c}	
\hline \hline 
Atom &  100\,K &  200\,K &  300\,K \\ 
\hline
Ce - 4Ni & 2.9752(6) & 2.9836(6) & 2.9921(6) \\  
\hspace{6mm} 4Bi & 3.3210(7) & 3.3267(7) & 3.3317(7) \\ 
\hspace{6mm} 4Bi & 3.4061(7) & 3.4196(7) & 3.4338(8) \\ 
\hline
Bi - 3Ni & 2.6992(6) & 2.7054(6) & 2.7114(6) \\  
\hspace{6mm} 3Ce & 3.3210(7) & 3.3267(7) & 3.3317(7) \\ 
\hspace{6mm} 3Ce & 3.4061(7) & 3.4196(7) & 3.4338(8) \\ 
\hspace{6mm} 3Bi & 3.6579(8) & 3.6717(8) & 3.6861(8) \\ 
\midrule
Ni - 4Bi & 2.6992(6) & 2.7054(6) & 2.7114(6) \\  
\hspace{6mm} 4Ce & 2.9752(6) & 2.9836(6) & 2.9921(6) \\  
\hline\hline
\end{tabular}
\label{table3}	
\end{table*}

\clearpage

\section{Relationship between stoichiometry, structural, and magnetic susceptibility data}\label{apx-stoichiometry}

\begin{figure*}[h]
\centering
\includegraphics[width=1\textwidth]{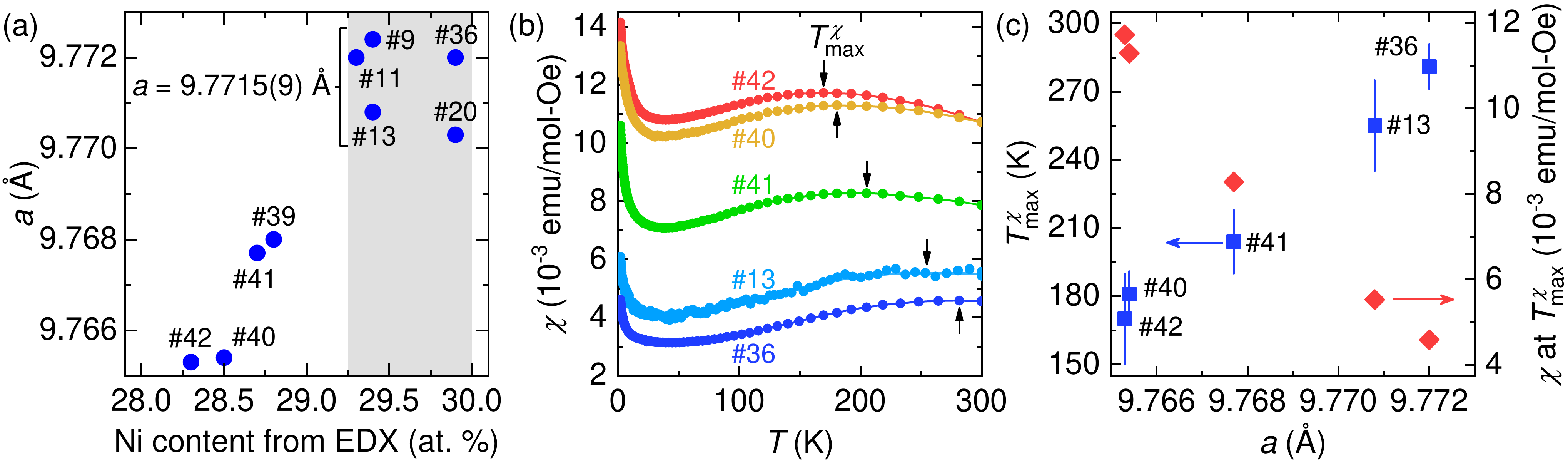}
\caption{(a) Lattice parameter $a$ vs Ni content of \cebini\ obtained from room-temperature powder XRD and EDX measurements, respectively, of selected samples from different growth batches (labeled by numbers). Data in the grey-shaded area correspond to the most stoichiometric samples (close to three Ni atoms per formula unit), with an average value of $a = 9.7715(9)$\,\AA. (b) Magnetic susceptibility $\chi$ vs $T$ of \cebini\ single crystals from different batches ($B \parallel$ [11-2] = 1\,T). Broad maxima at $T^{\chi}_{\mathrm{max}}$ are marked with arrows. (c) $T^{\chi}_{\mathrm{max}}$ (left scale) and susceptibility magnitude at $T^{\chi}_{\mathrm{max}}$ (right scale) obtained from (b) vs lattice parameter from (a).}
\label{fig-a-EDX-chi}
\end{figure*}

As discussed in Appendix \ref{apx-methods}, the growth of \cebini\ single crystals tends to yield samples with Ni deficiency. To evaluate how this affects the physical properties, we studied crystals from different batches, with Ni contents ranging from very close to 30\% (the exact 3:4:3 stoichiometry) to about 28.3\% (Fig.\,\hyperref[fig-a-EDX-chi]{\ref{fig-a-EDX-chi}}). Clear trends are observed. The lattice parameter $a$ increases with increasing Ni content (Fig.\,\hyperref[fig-a-EDX-chi]{\ref{fig-a-EDX-chi}(a)}). For the five most stoichiometric samples selected from different batches, shown in the gray-shaded area, the average lattice parameter is $a = 9.7715(9)$\,\AA.

Also the magnetic susceptibility $\chi$ changes systematically with the Ni content (Fig.\,\hyperref[fig-a-EDX-chi]{\ref{fig-a-EDX-chi}(b)}). The temperature of a local maximum seen in all $\chi(T)$ curves,  denoted by $T^{\chi}_{\mathrm{max}}$ (indicated with arrows), increases with increasing Ni content, whereas the magnitudes of $\chi$ at $T^{\chi}_{\mathrm{max}}$ decrease with increasing Ni content. Both trends are approximately linear (Fig.\,\hyperref[fig-a-EDX-chi]{\ref{fig-a-EDX-chi}(c)}).

This systematic behavior allows us to use magnetic susceptibility measurements to identify samples with a Ni content closest to the exact 3:4:3 stoichiometry. We have used samples from batch \texttt{\#}36 for the physical properties measurements presented in Sections \ref{results} and \ref{discussion}.

\section{Density functional theory calculations} \label{apx-DOS}

\begin{figure}[h]
\centering
\includegraphics[width=0.6\columnwidth]{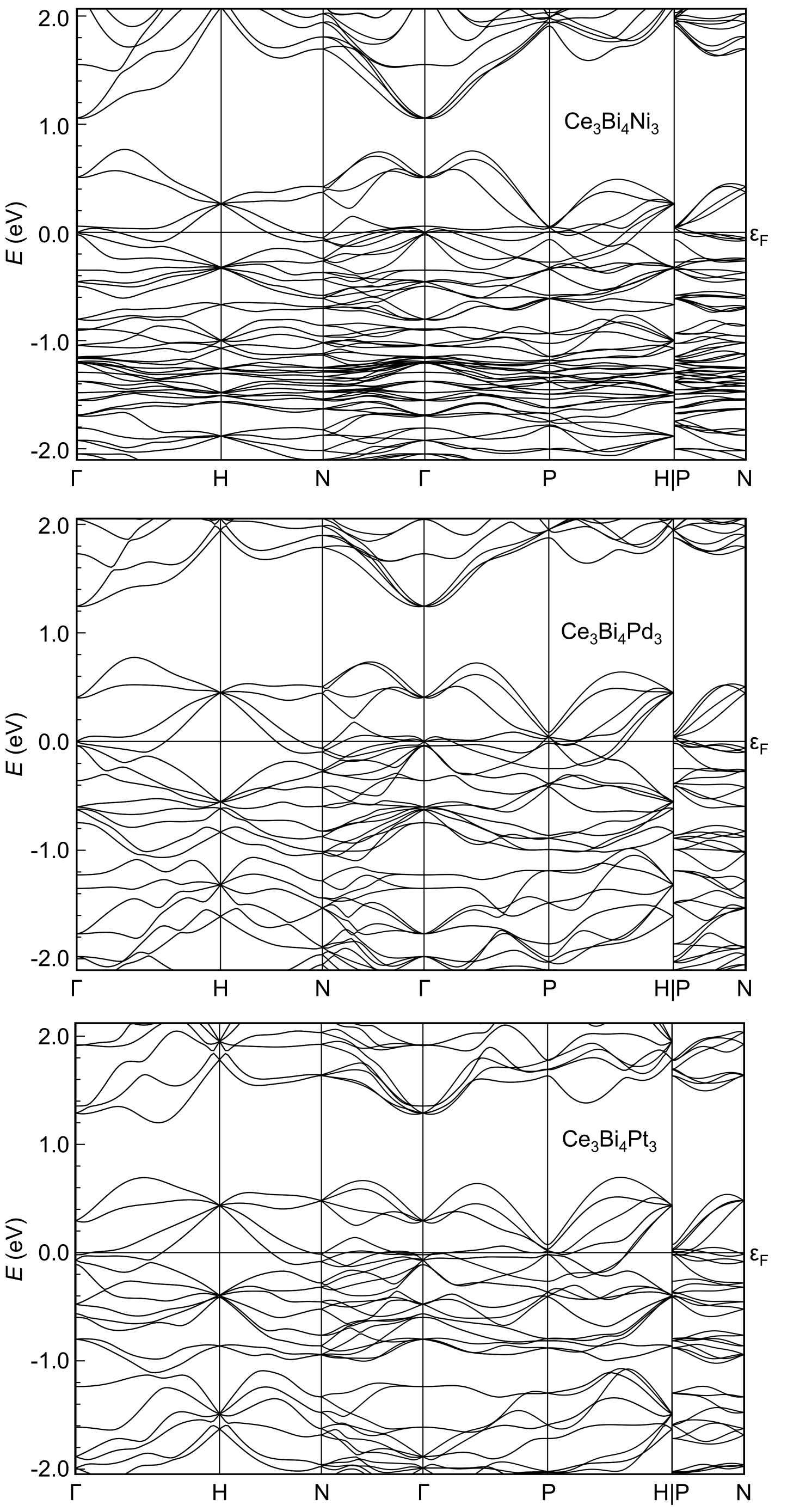}
\caption{Electronic structure of \cebix\ ($X =$ Ni, Pd, Pt), with the 4$f$ orbitals of Ce treated as core-like. The experimentally determined structure data were used as input. The Fermi energy $\varepsilon_{\mathrm{F}}$ is set to zero.}
\label{fig-DFT}
\end{figure}

The electronic structure of \cebix\ ($X =$ Ni, Pd, Pt) was calculated using FPLO-22 \cite{FPLO}, a density functional theory (DFT) package (Fig.\,\ref{fig-DFT}). Structural parameters obtained from the refinement of the single crystal XRD data were utilized (Appendix \ref{apx-structure}). The PBE (GGA) functional \cite{PBE} was used, as well as the default basis, with the 4$f$ orbitals of Ce treated as core-like. The latter means that no hybridization with $4f$ states is possible and the results thus represent the (hypothetical) situation without any Kondo interaction. As discussed in the main part, the motivation for these calculations is to evaluate whether the electronic structure evolves according to the atomic number of the elements ($3d$ element Ni: 28; $4d$ element Pd: 46; $5d$ element Pt: 78) or other effects are at work.

\begin{figure}[h]
\centering
\includegraphics[width=0.6\columnwidth]{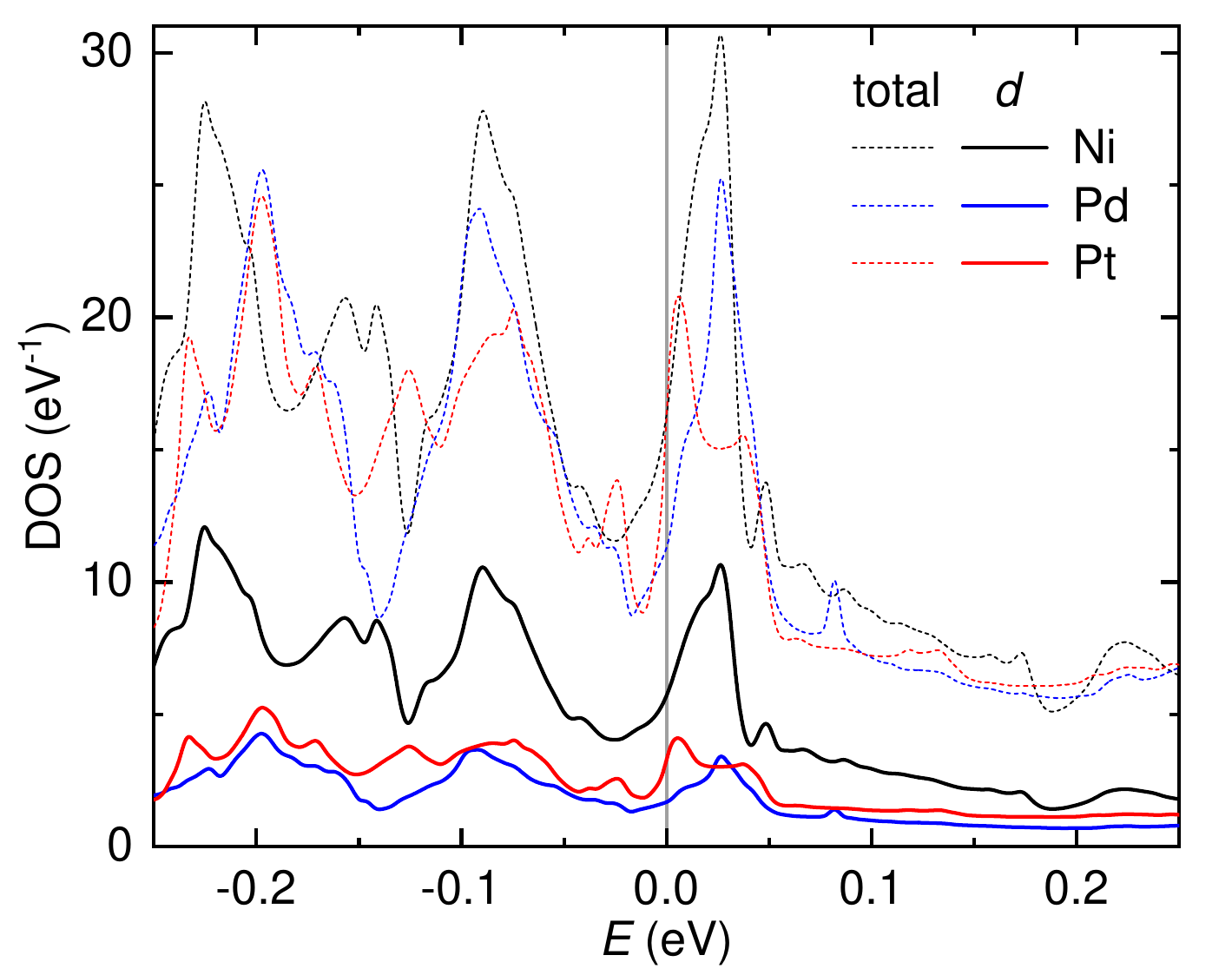}
\caption{Total and $d$-electron density of states ($d$\,DOS) of \cebix\ ($X= $\ Ni, Pd, Pt), obtained from the bandstructure calculations displayed in Fig.\,\ref{fig-DFT}. The Fermi energy is set to zero (vertical line).}
\label{fig-DOS}
\end{figure}

As the hybridization between $d$ and $f$ electrons is responsible for the experimentally observed Kondo effect, the density of states (DOS) of the former, $d$\,DOS, is the quantity of interest, in particular at the Fermi energy. We calculated it for all three compounds and the results are shown in Fig.\,\ref{fig-DOS}. Indeed, our calculations reveal that the DOS at the Fermi energy evolves nonmonotonically with the atomic number. It is largest for Ni, but larger for Pt than for Pd. The general expectation is that the smallest $3d$ orbitals have the largest DOS, and the largest $5d$ orbitals have the smallest DOS. However, the lattice parameters of the Pt and the Pd compound are essentially the same. This squeezes the $5d$ orbital of Pt, to yield the observed enhanced DOS.

Our DFT results have thus clarified the reason for the nonmonotonic trend in the physical properties across the Ni-Pd-Pt series of \cebix\ compounds.

\clearpage


\begin{thebibliography}{40}
\bibitem [{not()}]{note_ajeesh}%
  \BibitemOpen
  \href@noop {} {}\bibinfo {note} {In Ref.\,\cite{ajeesh22a}, the values of
  activated energy, which amount to half of the total energy gap $\Delta$ used
  here, were obtained from Arrhenius fits. For example, the authors reported an
  activated energy of 28\,K at 2.3\,GPa, which is equivalent to an energy gap
  of 56\,K $\sim$ 4.8\,meV.}\BibitemShut {Stop}%
\end{thebibliography}
\end{document}